\documentclass[twocolumn,prb,aps,showpacs,superscriptaddress, amsmath, amssymb,notitlepage]{revtex4-2}
\usepackage[dvipdfmx]{graphicx} 
\usepackage{dcolumn}
\usepackage{bm}
\usepackage{amsmath}
\usepackage{braket}
\usepackage{amsfonts}
\usepackage{amssymb}
\usepackage{color}
\usepackage{hyperref}
\hypersetup{colorlinks=true, citecolor= blue, linkcolor= blue, filecolor= blue, urlcolor= blue}
\usepackage{comment}
\usepackage{physics}
\usepackage {nonfloat}
\usepackage{multirow}
\usepackage{ulem}
\usepackage[dvipsnames]{xcolor}

\newcommand{\etal}{\textit{et al}.}
\newcommand{\bg}{\begin{pmatrix}}
\newcommand{\ed}{\end{pmatrix}}
\newcommand{\ep}{\epsilon}
\newcommand{\al}{\alpha}
\newcommand{\bt}{\beta}
\newcommand{\gm}{\gamma}

\begin{document}
\title{$Z_2$ flux binding to higher-spin impurities in the Kitaev spin liquid}
\author{Masahiro O. Takahashi}
\email{takahashi@blade.mp.es.osaka-u.ac.jp}
\affiliation{Department of Materials Engineering Science, Osaka University, Toyonaka 560-8531, Japan}
\author{Wen-Han Kao}
\email{kao00018@umn.edu}
\affiliation{School of Physics and Astronomy, University of Minnesota, Minneapolis, Minnesota 55455, USA}
\author{Satoshi Fujimoto}
\email{fuji@mp.es.osaka-u.ac.jp}
\affiliation{Department of Materials Engineering Science, Osaka University, Toyonaka 560-8531, Japan}
\author{Natalia B. Perkins}
\email{nperkins@umn.edu}
\affiliation{School of Physics and Astronomy, University of Minnesota, Minneapolis, Minnesota 55455, USA}

\begin{abstract}
Stabilizing $Z_2$ fluxes in Kitaev spin liquids (KSLs) is crucial for both characterizing candidate materials and identifying Ising anyons.
In this study, we investigate the effects of spin-$S$ magnetic impurities embedded in the spin-1/2 KSL.
Utilizing exact diagonalization and density matrix renormalization group methods,
we examine the impurity magnetization and ground-state flux sector with varying impurity coupling and spin size.
Our findings reveal that impurity magnetization exhibits an integer/half-integer spin dependence, which aligns with analytical predictions,
and a flux-sector transition from bound-flux to zero-flux occurs at low coupling strengths, independent of the impurity spin.
Notably, for spin-3/2 impurities, we observe a reentrant bound-flux sector, which remains stable under magnetic fields.
By considering fermionic representations of our spin Hamiltonian,
we provide phenomenological explanations for the transitions.
Our results suggest a novel way of binding a flux in KSLs, beyond the proposals of vacancies or Kondo impurities.
\end{abstract}

\maketitle
\section*{Introduction}
Magnetic impurities in strongly correlated electron systems are valuable tools for probing hidden physical phenomena.
A well-known example is the Kondo effect, where the scattering of conducting electrons by magnetic impurities in metals or quantum dots
reveals crucial insights into the low-energy physics of both the bulk material and the impurities \cite{Anderson1961, Kondo1964, Schrieffer1966, Gruner1974, Andrei1983}.
This effect, characterized by the screening of the magnetic impurity by the conduction electrons leading to the formation of a Kondo singlet,
has profound implications for understanding many-body interactions and has been extensively studied in both theoretical and experimental contexts.

The study of magnetic impurities extends to a variety of systems beyond conventional metals.
In topological insulators, for instance, magnetic impurities can break time-reversal symmetry,
leading to the opening of a gap at the Dirac point on the surface states and potentially inducing novel magnetic phases \cite{Liu2009, Biswas2010}.
These systems provide a rich playground for exploring the interplay between magnetism and topology, with potential applications in spintronics and quantum computing.

In low-dimensional spin systems such as quantum spin liquids (QSLs), the introduction of impurities can reveal even more exotic phenomena \cite{Kolezhuk2006, Chen2020, He2022, Lee2023}.
QSLs, which are characterized by a lack of conventional magnetic order even at zero temperature due to strong quantum fluctuations,
offer a unique environment where impurities can induce localized excitations and modify the emergent gauge fields.

In the context of the Kitaev spin liquid (KSL) model \cite{Kitaev2006}
— a paradigmatic example of a two-dimensional QSL with fractionalized excitations and emergent gauge fields —
introducing impurities, whether magnetic (spin-$S$ sites) or non-magnetic (vacancies)
\cite{Willans2010, Willans2011, Santhosh2012, Das2016, Vojta2016, Kao2021vacancy, Kao2021localization, Dantas2022, Takahashi2023, Vlad2024},
can lead to various novel phenomena.
These include localized bound states \cite{Santhosh2012, Kao2021localization, KaoPRL2024, KaoPRB2024},
flux binding effects \cite{Willans2010, Willans2011, Vojta2016, Kao2021vacancy, Vlad2024},
and modifications in the system's topological nature \cite{Dantas2022, Vlad2024}. 
Such effects not only provide new insights into impurity physics in QSLs but also enhance our understanding of their overall behavior. 
Due to the localized nature of the low-energy fractionalized excitations in the above phenomena, characterization of KSL based on the signatures in scanning tunneling microscopy (STM) has been proposed in various theoretical works \cite{Knolle2020, Pereira2020, Konig2020, Udagawa2021, Bauer2023, Takahashi2023, KaoPRL2024, KaoPRB2024, Bauer2024, Jahin2024, Zhang2024short, Zhang2024long}.
In particular, the above proposals show that the inelastic tunneling spectroscopy can access the real-space spin-spin correlation function of the Kitaev model, which is very sensitive to defects, open edges, and local flux structures.

So far, two types of local impurities have been relatively well-studied in the Kitaev model.
The first type is vacancies.
It has been demonstrated that vacancies in the Kitaev model lead to almost zero-energy localized bound states and flux-binding effects \cite{Willans2010,  Willans2011, Santhosh2012, Zschocke2015,  Kao2021vacancy, Kao2021localization},
which can potentially be probed by thermodynamics \cite{Kitagawa2018spin, Imamura2024} and STM \cite{Takahashi2023, KaoPRL2024, KaoPRB2024}.
The second type is spin-$S$ impurities, which are coupled to KSL at a given site via Kondo coupling.
The studies of Kondo impurities \cite{Dhochak2010, Das2016, Vojta2016} have highlighted several remarkable properties of the Kondo effect in the Kitaev model.
In the presence of a spin-1/2 Kondo impurity, the fluxes in the three plaquettes adjacent to the impurity site are no longer individually conserved.
However, their product (the flux in the impurity plaquette) and all outer fluxes remain conserved \cite{Vojta2016}.
Furthermore, a topological transition occurs from the zero-flux state to a bound-flux state attached to the impurity site as a function of Kondo coupling \cite{Das2016, Vojta2016}.

In this work, we theoretically investigate the behavior of $S_{\rm{imp}}=1$ and $S_{\rm{imp}}=3/2$ impurities in KSL
by means of numerical exact diagonalization (ED) and density matrix renormalization group (DMRG) methods as well as phenomenological models in the Majorana representation.
We will focus on the case with a single magnetic impurity,
although the case of multiple impurities can be straightforwardly extended.

We show that the behavior of KSL with a magnetic impurity
strongly depends on whether the impurity has a half-integer or integer spin.
This dependence, which we demonstrate by considering two cases of magnetic impurities,
$S_{\rm{imp}}=1$ and $S_{\rm{imp}}=3/2$, echoes the recent findings by Ma \cite{Ma2023},
which shows that the nature of the spin-$S$ $Z_2$  KSL differs based on whether the spin is integer or half-integer.  While Ma's work introduces $2S$-flavor Majorana representation for the pure spin-$S$ KSL and identifies the $Z_2$ gauge fluxes as conserved quantities, we investigate the behavior of a spin-$S$ impurity embedded within a spin-1/2 Kitaev spin liquid, focusing on the impurity magnetization and
the ground-state flux sector as functions of varying impurity coupling strength and spin size. This mixed-spin KSL system, though less explored in the literature, holds significant potential for realization in Kitaev materials containing magnetic impurities.

Similarly to vacancies, the magnetic impurities can bind $Z_2$-fluxes in the lattice.
We show that varying the coupling of spin-$S$ impurity with the surrounding spin-1/2 KSL can drive a phase transition between bound-flux and zero-flux sectors.
Furthermore, the point at which this transition occurs depends on the magnitude of $S_{\rm{imp}}$.
This is significant because, in the presence of a time-reversal symmetry-breaking magnetic field,
each $Z_2$ flux can bind a Majorana zero mode,
thereby realizing an Ising anyon governed by non-Abelian statistics \cite{Kitaev2006}.
Therefore, demonstrating that magnetic impurities can trap $Z_2$ fluxes
provides a pathway to realizing Ising anyons in these systems.

\section*{Results}
\begin{figure*}[t!]
    \centering
    \includegraphics[width=170mm]{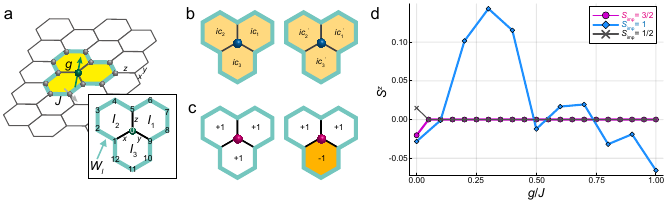}
    \caption{ Integer/half-integer dependence of magnetic impurity.
    \textbf{a}  A spin-$S$ impurity embedded in the Kitaev spin liquid. The inset shows site labels around the impurity site (0) and the plaquette labels used in Eq.~\ref{Eq:impurity_plaquette}.
       \textbf{b} and  \textbf{c}: Typical expectation values of internal flux operators ${\mathbb{W}}_a$ including the impurity site (0) in the bound-flux sector (left) and the zero-flux sector (right) for $S_{\rm{imp}}=1$ and $S_{\rm{imp}}=3/2$, respectively.
    \textbf{d} Local magnetization of the impurity spin embedded in spin-1/2 Kitaev spin liquid, calculated via spin-space exact diagonalization in a 24-site cluster.}
    \label{fig:local_spin}
\end{figure*}
\subsection*{Model}
To describe the Kitaev model with magnetic impurities, we begin with the Hamiltonian:
\begin{align}
    H&=-J\sum_{\begin{subarray}{c} j,k\notin\Lambda, \\ \langle jk\rangle_\mu\end{subarray}}{S}_j^\mu {S}_k^\mu
    -g\sum_{\begin{subarray}{c} {j}\in\Lambda, k\notin\Lambda,\\ \langle {j}k\rangle_\mu\end{subarray}}
    {{\tilde{S}}}_{j}^\mu {S}_{k}^\mu\quad (\mu=x,y,z).
    \label{eq:Ham}
\end{align}
Here, ${S}^\mu=\sigma^\mu/2$ represents the $\mu$-component of the spin-1/2 operator, with $\bm{\sigma}$ denoting the Pauli matrices.  
${\tilde{S}}^\mu$ represents the $\mu$-component of the spin-$S$ operator of the impurity. The set of impurity sites is denoted by $\Lambda$. 
$J>0$ denotes the ferromagnetic coupling strength between spin-1/2 operators away from the impurity,
while $g>0$ denotes the ferromagnetic coupling strength between the impurity spins and the spin-1/2 operators of the original Kitaev model.
 We mainly consider the case that a single impurity is not located on the edges.

Since the interaction on each honeycomb bond remains Kitaev-like, we can define a triple-plaquette flux operator in the vicinity of the magnetic impurity as:
\begin{align}
    W_I\equiv 2^{12} S_1^x S_2^x S_3^y S_4^z S_5^z S_6^z S_7^x S_8^y S_9^y S_{10}^y S_{11}^z S_{12}^x,
    \label{eq:W_I}
\end{align}
where the product is taken over all bonds forming a 12-site plaquette
around the impurity site labeled as ``0'' (see Fig. \ref{fig:local_spin}\textbf{a}).
This operator captures the flux configuration within the three plaquettes surrounding the magnetic impurity and allows us to analyze the effect of the impurity on the local flux dynamics.
A $Z_2$ flux operator at a plaquette $p$ in the bulk,
where there is no impurity spin, is symbolically expressed as
${W}_p=2^6 \prod_{j\in p}S_{j}^\mu$ in the spin-1/2 basis \cite{Kitaev2006}.
Both ${W}_p$ and ${W}_I$ commute with the Hamiltonian ${H}$ and with each other, taking $\pm1$ as their eigenvalues, respectively.
Consequently, the total Hilbert space is divided into individual flux sector subspaces:
$\mathcal{L}=\bigoplus_{w_{p_1},w_{p_2},\cdots,w_I}\,\mathcal{L}_{w_{p_1},w_{p_2},\cdots,w_I}$.
We refer to the sector with $w_I=-1$ as the bound-flux sector
and the sector with $w_I=+1$ as the zero-flux sector,
when all plaquettes not in the vicinity of the impurity have $w_p=1$, as introduced in a vacancy case \cite{Willans2010}.

\subsection*{Internal plaquette operators}
Around the impurity, the three adjacent plaquette operators must incorporate the higher-spin operator $\tilde{S}^{\alpha}_0$,
such that they are constructed by the unitary operators of $\pi$-rotation, as introduced by Baskaran \etal \cite{Baskaran2008}.
We first quote the basic properties of these unitary operators \cite{Tasaki2020}:
\begin{align}
\begin{split}
\tilde{R}^{\al}_{j} = e^{i\pi \tilde{S}^{\al}_{j}}, \quad \left(\tilde{R}^{\al}_{j}\right)^2 = (-1)^{2\tilde{S}_{j}},\\ \tilde{R}^{\al}_{j} \tilde{R}^{\bt}_{j} = (-1)^{2\tilde{S}_{j}}\tilde{R}^{\bt}_{j}\tilde{R}^{\al}_{j}, \quad \tilde{R}^{\al}_{j} \tilde{R}^{\bt}_{j} = \tilde{R}^{\gm}_{j},
\end{split}
\end{align} 
where $\al \neq \bt \neq \gm$, and $(\al,\bt,\gm)\in(x,y,z)$ obey cyclic permutations.
The above relations are defined on the same site ${j}$,
otherwise the operators simply commute on different sites.
For a general hexagonal plaquette with arbitrary spin on the corners,
one can define the plaquette operator as
\begin{align}\label{Wp}
\mathbb{W}_p = \prod_{{j} \in p}\tilde{R}^{\al_{j}}_{j}, \quad (\mathbb{W}_p)^2 = \prod_{{j} \in p}(-1)^{2\tilde{S}_{j}}.
\end{align}

Specifically, for the three plaquettes shown in Fig.~\ref{fig:local_spin}(a), we can write down  (the tilde is dropped for spin-1/2 sites for clarity)
\begin{align}\label{Eq:impurity_plaquette}
\begin{split}
&\mathbb{W}_{I_1} = R^z_6\,R^x_7\,R^y_8\,R^z_9\,\tilde{R}^x_{0}\,R^y_5,\\
&\mathbb{W}_{I_2} = R^z_4\,R^x_5\,\tilde{R}^y_{0}\,R^z_1\,R^x_2\,R^y_3,\\
&\mathbb{W}_{I_3} = \tilde{R}^z_{0}\,R^x_9\,R^y_{10}\,R^z_{11}\,R^x_{12}\,R^y_1,
\end{split}
\end{align}
and then it is straightforward to show that $\mathbb{W}_{I_1}\mathbb{W}_{I_2} = (-1)^{2S_{\rm{imp}}+2S_5}\mathbb{W}_{I_2}\mathbb{W}_{I_1}$.
Therefore, one can see that the mutual commutation relation between the internal plaquette operators depends on the size of the spins shared by the two hexagons,
which is consistent with the previous work on the mixed-spin Kitaev model~\cite{Koga2019}.
In our impurity model, only the impurity site can have a general spin size and all the rest are spin-1/2.
This leads to the important properties of the internal plaquette operators:
\begin{align}
\begin{cases}
\left[{\mathbb{W}}_{a},\,{\mathbb{W}}_{b}\right]=0, \quad (\mathbb{W}_a)^2 = +1, & $if$~S_{\rm{imp}}~$is a half-integer$,  \\
\left\{{\mathbb{W}}_{a},\,{\mathbb{W}}_{b}\right\}=0,\,\,\, (\mathbb{W}_a)^2 = -1, & $if$~S_{\rm{imp}}~$is an integer$,
\end{cases}
\label{eq:size_dependent_relation}
\end{align}
where $a\neq b$ and $(a,b)\in (I_1,I_2,I_3)$.

Note that a triple-plaquette operator and internal plaquette operators are directly related. They satisfy the following relation:
\begin{align}
{\mathbb{W}}_{I_1}{\mathbb{W}}_{I_2}{\mathbb{W}}_{I_3}=(-1)^{2S_{\rm{imp}}}{W}_I.
\end{align}
For a half-integer spin impurity, the conserved value $w_I(=\pm1)$ can be decomposed into the product of $w_{I_1},w_{I_2},$ and $w_{I_3}$, each taking $\pm1$.
Thus, in the case of the $S_{\rm{imp}}=3/2$, the triple-plaquette operator ${W}_I$ and three internal plaquette operators ${\mathbb{W}}_{a}$ are related by ${W}_I = -{\mathbb{W}}_{I_1}{\mathbb{W}}_{I_2}{\mathbb{W}}_{I_3}$.
There are four possible configurations for $(w_{I_1},w_{I_2},w_{I_3})= (+1,+1,+1), (+1, -1, -1), (-1, +1, -1)$ and $(-1,-1,+1)$ that can realize the bound-flux sector $w_I=-1$ (see the left panel of 
Fig. \ref{fig:local_spin}\textbf{c}).
Similarly, there are four degenerate internal flux configurations that can realize the zero-flux sector (see the right panel of Fig. \ref{fig:local_spin}\textbf{c}).
In contrast, for $S_{\rm{imp}}=1$, the internal ${\mathbb{W}}_{a}$ always takes on a purely imaginary value with some real coefficient $c\in\mathbb{R}$.
This makes it independent of ${w}_I$ (see Fig. \ref{fig:local_spin}\textbf{b}). These distinct behaviors for different impurity spin values are clearly observed in the DMRG calculations, which we will discuss more later.

\subsection*{Integer/half-integer dependence of the impurity magnetization}
Based on the algebra of the internal plaquette operators \eqref{eq:size_dependent_relation},
one can derive an interesting integer/half-integer effect on the magnetization of the impurity spin. Notice that each internal plaquette operator contains only one of the three spin components of the impurity.
This leads to the commutation/anticommutation relations between the internal plaquette operators and impurity spin operators: 
\begin{align}
\begin{split}
&\left[\mathbb{W}_{I_1}, \tilde{S}^x_0\right] = \left\{\mathbb{W}_{I_1},\tilde{S}^y_0\right\} =\left\{\mathbb{W}_{I_1},\tilde{S}^z_0\right\} = 0\\
&\left[\mathbb{W}_{I_2}, \tilde{S}^y_0\right] = \left\{\mathbb{W}_{I_2},\tilde{S}^z_0\right\} =\left\{\mathbb{W}_{I_2},\tilde{S}^x_0\right\} = 0\\
&\left[\mathbb{W}_{I_3}, \tilde{S}^z_0\right] = \left\{\mathbb{W}_{I_3},\tilde{S}^x_0\right\} =\left\{\mathbb{W}_{I_3},\tilde{S}^y_0\right\} = 0.
\end{split}
\end{align}
From Eq. \eqref{Wp}, it follows that if the $S_{\rm{imp}}$ is a half-integer,
the square of the internal plaquette operators is one ($\mathbb{W}_a^2 = +1$)
and they all mutually commute ($\left[\mathbb{W}_a,\mathbb{W}_b\right] = 0$). 
Using these properties, we can demonstrate that:
\begin{align}
\expval{\tilde{S}^z_0} &= \expval{\tilde{S}^z_0\,\mathbb{W}_{I_1}^2} \nonumber\\
&= -\expval{\mathbb{W}_{I_1}\,\tilde{S}^z_0\,\mathbb{W}_{I_1}} = -(w_{I_1})^2\expval{\tilde{S}^z_0} = -\expval{\tilde{S}^z_0},
\end{align}
which implies $\expval{\tilde{S}^z_0} = 0$.
The derivation is applicable to the other two components,
$\expval{\tilde{S}^x_0} = \expval{\tilde{S}^y_0} = 0$,
because one can always find an internal plaquette operator that anticommutes with the impurity spin component.
The same argument applies to the three neighboring components ($S^x_1$, $S^y_9$, and $S^z_5$),
as well as to all other spin-1/2 operators on the lattice.
In contrast, for an integer-spin impurity, $w_{I_a}$ takes pure imaginary values, not quantized to $Z_2$, 
leading to the possibility of a non-zero local magnetization $\langle\tilde{S}_{0}^z\rangle\neq0$.
Similarly, the $\mu$-component ($\mu = x,y,z$) of spin-1/2 neighboring with the impurity spin on $\mu$-bond can have a nonzero spin moment; the other two components should be zero.

This spin-size dependence of the impurity magnetization can be confirmed through numerical exact diagonalization, as illustrated in Fig. \ref{fig:local_spin}\textbf{d}.
Here we perform ED calculation for the spin Hamiltonian in a 24-site cluster on a cylinder geometry with a single impurity to evaluate the $z$-component of the magnetic impurity's local spin moment. 
For any finite $g$, the local spin moment is numerically zero for both $S_{\rm{imp}}=1/2$ and 3/2 cases, which aligns precisely with our analytical findings. 
Conversely, in the $S_{\rm{imp}}=1$ case, the local moment takes finite values in the range $[-1, 1]$ at random due to local dynamics induced by the anticommutation relations of internal plaquette operators.
The points plotted in Fig. \ref{fig:local_spin}\textbf{d} represent the results obtained in a single run.
Averaging $\tilde{S}_0^z$ across multiple independent runs would yield a mean value approaching zero, highlighting a clear qualitative difference from the half-integer impurity case.
\begin{figure*}
    \centering
    \includegraphics[width=170mm]{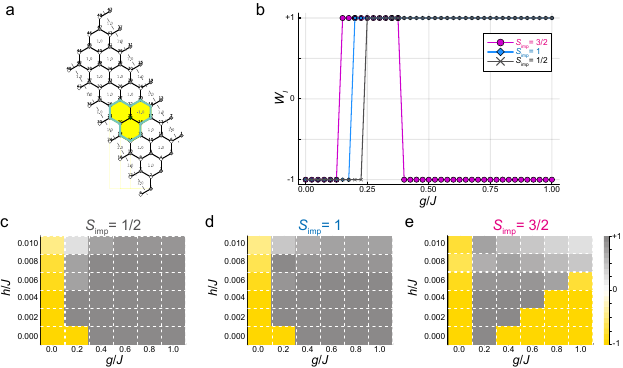}
    \caption{DMRG results of flux-sector transitions.
    \textbf{a} The bound-flux sector of a 48-site cluster used in calculations.
    Numbers of every plaquette in the bulk shows $\langle {W}_p\rangle$, while the number next to the impurity site shows $\langle {W}_I\rangle$.
     \textbf{b}  Flux-sector transition calculated by DMRG with a single spin-$S$ impurity embedded in spin-1/2 Kitaev spin liquid. \textbf{c}, \textbf{d} and \textbf{e}: Phase diagrams of $\langle {W}_I\rangle$ 
     as a function of the magnetic field strength $h/J$
     and the impurity-bulk coupling strength $g/J$
     for the cases $S_{\rm{imp}}=1/2$, $1$, and $3/2$, respectively.
    We use the same color scale for all three subfigures: the yellow (gray) region corresponds to the bound-flux (zero-flux) sector.
    White dot lines denote the edges of each pixel.
  }
    \label{fig:DMRG_results}
\end{figure*}

\subsection*{Reentrance effect of the bound-flux sector in zero and finite magnetic fields}
Here, we focus on the ground-state flux configuration in the presence of the impurity.
Using DMRG, we numerically determine the ground-state flux configurations as a function of
$g/J$ in a 48-site cylinder with a single spin-$S$ impurity located in the bulk
[see Fig. \ref{fig:DMRG_results}\textbf{a}].
Fig. \ref{fig:DMRG_results}\textbf{b} shows the transitions between the bound-flux and zero-flux sectors.
Starting from the bound-flux sector at $g=0$,
the system undergoes a flux-sector transition under a weak but nonzero coupling $g$,
leading to the zero-flux sector.
This behavior is qualitatively consistent across all $S_{\rm{imp}}$ cases,
indicating the instability of the bound-flux sector
in a ``quasivacancy'' problem as reported in  Ref. \cite{Zschocke2015, Kao2021vacancy}.
However, the transition point $g_1$ depends on the spin size of the impurity, shifting to the left as $S_{\rm{imp}}$ increases.
Additionally, for the $S_{\rm{imp}}=3/2$ impurity case, the system undergoes a second flux-sector transition at the transition point $g_2$, resulting in the appearance of the reentrant bound-flux sector.
Note again that, in both flux sectors of the $S_{\rm{imp}}=3/2$ impurity case, a triple-plaquette operator ${W}_I$ and three internal plaquette operators ${\mathbb{W}}_{a}$ are related as ${W}_I = -{\mathbb{W}}_{I_1}{\mathbb{W}}_{I_2}{\mathbb{W}}_{I_3}$.
Thus, each flux sector is fourfold degenerate.

A few remarks are in order.
 First, the flux gap exhibits a position dependence of the impurity, possibly due to the edge effect on the flux gap as described by Feng {\it et al.} \cite{Feng2020}. Specifically, the flux gap decreases in the cylinder geometry as the impurity approaches one of the edges.
This positional dependence might contribute to the energy difference between the two flux sectors.
For $S_{\rm{imp}}=3/2$, this effect is mild and
only changes the transition points $g_1$ and $g_2$.
In contrast, for $S_{\rm{imp}}=1$,
we observe a significant qualitative change in the flux-sector transitions in the strong $g$ region
due to the position of the impurity, even on the 48-site cylinder.
This change is influenced not only by the specifics of the numerical conditions — such as cluster shape, size, and boundary conditions, — but also by the dynamical properties around the impurity, particularly the lack of quantization of internal flux operators ${\mathbb{W}}_a$, as described at 
Eq.~\eqref{eq:size_dependent_relation}.
 Second, the two-impurity case exhibits behavior qualitatively similar to the single-impurity case. Specifically, every impurity spin with $S_{\rm{imp}} = 3/2$ binds the $Z_2$ flux in a wide parameter range, while for $S_{\rm{imp}} = 1$, there is a strong position dependence on the flux-sector transitions.
Third, our main findings are not specific for the 48-site cylindrical cluster. We have confirmed the same trends in other finite-size cylindrical clusters with different shapes.
Numerical evidence supporting these arguments can be found in the Supplementary information \cite{SM}.

We also examined the stability of the bound-flux sector under a uniform magnetic field numerically using DMRG in the same finite-size cluster.
For this analysis, we considered the Hamiltonian
$H_{\rm{total}} = {H} + h \sum_{j,\mu} S_j^\mu$,
with the field applied in the [111] direction in the spin basis.
We obtained three phase diagrams of $\langle W_I \rangle$ as a function of $g$ and $h$,
shown in Figs. \ref{fig:DMRG_results}\textbf{c}, \textbf{d}, and \textbf{e},
corresponding to the cases of $S_{\rm{imp}}=$1/2, 1, and 3/2, respectively.

In the cases of $S_{\rm{imp}}=$1/2 and 1, the bound-flux sector at nonzero $g$ is very fragile
in the presence of the external field, and this fragility is independent of the impurity's position \cite{SM}.
In contrast, the reentrant bound-flux sector for a spin-3/2 impurity exhibits some stability for finite field strengths.
This stability also depends on the coupling strength $g$:
as $g$ increases, the bound-flux sector tends to withstand stronger fields.
This behavior can be understood by noting that the energy difference
between the bound-flux and zero-flux sectors under zero magnetic field increases monotonically with $g/J$ in the reentrant flux sector regime.
This increase in energy difference provides the bound-flux sector with greater stability in the presence of magnetic fields.

\subsection*{Triple-plaquette analysis: Majorana representation and effective-coupling model}
In this section, we aim to understand better the previous numerical findings, including the spin-size-dependent flux-sector transitions and the reentrance of the bound-flux sector.
Similar phenomena have been studied in the site-diluted KSL, where a $\pi$-flux can be trapped by a true vacancy or quasivacancy \cite{Willans2010, Zschocke2015, Takahashi2023, Kao2021vacancy, KaoPRB2024}. 
These studies were based on the exact solution of  KSL using the Majorana representation for spin-1/2 \cite{Kitaev2006}.
In these studies, some of the authors have shown that the low-energy modes introduced by quasivacancies are highly localized, allowing for a clear distinction in the energy spectrum between zero- and bound-flux sectors \cite{Kao2021vacancy, KaoPRB2024}.
Recently,  on the other hand, the Majorana parton construction was generalized to study uniform spin-$S$ KSLs \cite{Ma2023}.
The essential idea of this construction is recognizing spin-$S$ operator as $2S$ flavors of spin-1/2s with constraints.
This motivates us to re-examine our findings in the mixed-spin KSL
using the fermionic approach  and the multiple-flavor representation, which we outlined in the  Methods.
To this end, we focus on a triple-plaquette  cluster, 
as shown in the inset of Fig. \ref{fig:local_spin}\textbf{a},
to demonstrate that even the system on the minimal cluster can capture the key findings from the previous section.

In the following, we will discuss 
 two fermionic models in the triple-plaquette cluster:
the $S_{\rm{imp}} = 3/2$ impurity model with four-body Majorana interaction term,
and the effective-coupling model for a general $S_{\rm{imp}}$.
In a general case, 
the  Hamiltonian \eqref{eq:Ham}
with a magnetic impurity in Majorana fermion representation reads:
\begin{align}
    H&=\frac{J}{4}\sum_{\begin{subarray}{c} j,k\notin {\Lambda}, \\ \langle jk\rangle_\mu\end{subarray}}\,u_{jk}^\mu\,ic_jc_k
    +\frac{g}{4}\sum_{\begin{subarray}{c} j\in{\Lambda},k\notin {\Lambda}, \\ \langle jk\rangle_\mu\end{subarray}}\sum_{a=1}^{2\tilde{S}}
    (i\gamma_{aj}^\mu b_k^\mu)(i\gamma_{aj}^0 c_k),\label{HamMajorana}
\end{align}
where $\Lambda=0$ denotes the impurity site in the triple-plaquette  cluster (see, again, the inset of Fig. \ref{fig:local_spin}\textbf{a}).
Here, we introduce $b^\mu$ and $c$ Majorana operators for the bulk spin-1/2 operator
and $\gamma$-Majorana operators for the impurity spin  as used in Ref. \cite{Ma2023}.
The $Z_2$ gauge field on all the 12 edges is conserved and has eigenvalues $u_{jk}^\mu = \pm 1$.
The triple-plaquette flux, $w_I$, is thus determined by $w_I = \prod u_{jk}^\mu$. 
The second term in Eq. \eqref{HamMajorana} represents the transformed impurity coupling, which becomes a four-Majorana interaction.
Since $i\gamma_{aj}^\mu b_k^\mu$ does not commute with $H$,
this quartic term can not be trivially rewritten as a quadratic form.
In our  ED calculation, this four-Majorana interaction is treated as it is
by preparing $2M$ Majorana fermion operators composed of $M(\in \mathbb{N})$ complex fermion operators in a binary number basis.

\begin{figure*}
    \centering
    \includegraphics[width=170mm]{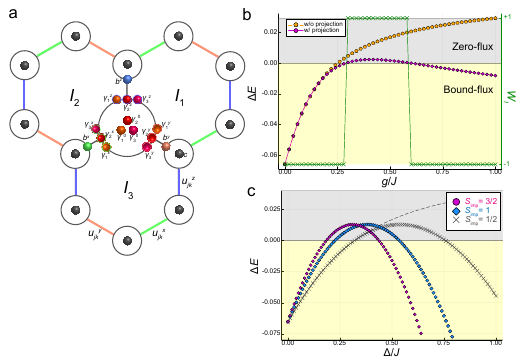}
    \caption{ Fermionic representation and analysis for the triple-plaquette cluster.
    \textbf{a}  
    Majorana representation  for the spin-3/2 impurity case in the triple-plaquette cluster, following the parton construction by Ma \cite{Ma2023}. The remaining sites 
    are spin-1/2 degrees of freedom, described using Kitaev's original parton construction  \cite{Kitaev2006}.
    \textbf{b} Energy difference, $\Delta E$,  between the bound-flux and zero-flux sectors for the $S_{\rm{imp}} = 3/2$ case. Magenta (orange) curve: $\Delta E$ computed with (without) projection operators. Green line: $ {w}_I$  calculated in the spin Hamiltonian \eqref{eq:Ham}.
 \textbf{c}  $\Delta E$ calculated in the effective-coupling model, Eq \eqref{eq:quadratic}.
 The coefficients are given in the main text. From $S_{\rm{imp}} = 1/2$ to higher impurity spins,
 the substitution $\Delta \to \sqrt{2S_{\rm{imp}}}\Delta$ is applied.}
    \label{fig:Majorana_rep}
\end{figure*}

We visualize in Fig. \ref{fig:Majorana_rep}\textbf{a} the triple-plaquette cluster with $S_{\rm{imp}} = 3/2$ in the Majorana representation.
In this case, there are 27 Majorana fermions in the whole 13-spin-site system,
including 12 from $c$-Majoranas of spin-1/2s, 12 from $\gamma$-Majoranas of the impurity site, and 3 from $b$-Majoranas at nearest-neighbor sites of the impurity. 

Using ED for the many-body Majorana Hamiltonian \eqref{HamMajorana} under projections for the spin-3/2 impurity,
we calculate the energy difference between the bound-flux ($w_I=-1$) and zero-flux ($w_I=1$) states of the triple-plaquette  cluster.
This energy difference, defined as
\begin{align}\label{eq:energy_diff}
\Delta E\equiv E_{\rm{bound}} - E_{\rm{zero}},
\end{align}
is evaluated as a function of $g/J$ (see the magenta curve in Fig. \ref{fig:Majorana_rep}\textbf{b}).
When $\Delta E<0$ ($\Delta E>0$), the system realizes the bound-flux (zero-flux) sector at a given $g/J$, which is illustrated as a yellow (gray) shaded area in the figure.
We observe two key behaviors:
{\it i)} the initial transition from the bound-flux sector to the zero-flux sector at the transition point $g_1$,
and {\it ii)} the reentrant transition back to the bound-flux sector at the second transition point $g_2$.
For $g>g_2$, $\Delta E$ decreases monotonically, indicating the stability of the reentrant bound-flux sector.
Moreover, these two transition points match exactly with the flux-sector transition points obtained in ED of the spin-basis Hamiltonian.
The calculated $w_I$ in spin-basis ED is shown as the green line in the Fig. \ref{fig:Majorana_rep}\textbf{b}.
It is also worth mentioning here that the correct behavior of flux-sector transition calculated in the Majorana representation is only obtained under the presence of projection operators $P_a$ and $P_{\tilde{S}=3/2}$ (see Methods for details).
Without these projection operators, as shown by the orange curve in
Fig. \ref{fig:Majorana_rep}\textbf{b}), the reentrant bound-flux sector does not appear, emphasizing the need for proper constraints
on the four-body Majorana terms.

The presented numerical results confirm that even a minimal  cluster with a $S_{\rm{imp}} = 3/2$ impurity can qualitatively capture the flux-sector transitions and the reentrant effect,
provided a proper treatment of projection.
It reinforces the localized picture of the flux-binding effect by a site defect. 

To further extract the essential ingredient that leads to a reentrant bound-flux sector,
we propose a heuristic approach based on the effective coupling between the $c$-Majorana eigenmodes of the 12-site plaquette and the Majorana zero modes ($\gamma_{a}^0$) introduced by the impurity.
In the first constituent, the $L = 12$ plaquette is considered as a fermion hopping problem on a ring, where $L$ denotes the number of sites on the ring \cite{Moessner_Moore_2021}. 
The difference between the zero- and $\pi$-flux sectors is translated into the difference between the periodic and antiperiodic boundary conditions (PBC and APBC) of the ring (see Fig. \ref{fig:periodic_ring}\textbf{a}).
Here, the $\pi$-flux sector corresponds to the bound-flux sector we have discussed.
If the nearest-neighbor hopping strength is $J/4$ (corresponds to the Kitaev coupling in the spin Hamiltonian),
the free-fermion Hamiltonian on the ring reads:
\begin{align}
\mathcal{H}^{\rm{ring}}_{\rm{PBC/APBC}} = \frac{J}{4}\sum_{j = 1}^{L-1}\left(a^{\dagger}_j a_{j+1}+h.\,c.\right) \pm \frac{J}{4}\left(a^{\dagger}_{L}a_{1}+h.\,c.\right),
\end{align}
with the corresponding energy eigenvalues
\begin{align}\label{eq:epsilon_ring}
\begin{cases}
\epsilon_{\mathrm{PBC}} = -\frac{J}{2}\cos\left(\frac{2\pi n}{L}\right),\\
\epsilon_{\mathrm{APBC}} = -\frac{J}{2}\cos\left[\frac{(2n+1)\pi}{L}\right],
\end{cases}
\end{align}
where $n$ is the integers from $0$ to $L-1$.
Since the ground-state energy of the system is calculated by the sum of all negative-energy modes,
it is straightforward to conclude that $L = 4n+2$ favors the zero-flux sector (periodic boundary)
and $L = 4n$ favors the $\pi$-flux sector (antiperiodic boundary).
This is consistent with the prediction by Lieb's theorem of flux configuration \cite{Lieb1994}, even only a single plaquette being considered here \cite{Lieb1992}.
Note that in this $L = 4n$ case, $\epsilon_{\mathrm{PBC}}$ contains two zero modes
while $\epsilon_{\mathrm{APBC}}$ contains no zero modes under zero magnetic fields.
This is the key difference that helps us understand the flux-sector transition in the presence of impurity.

To tackle the quasivacancy or impurity problem with $S_{\rm{imp}} = 1/2$,
we add an additional fermion that only directly couples to three sites of the $L = 12$ ring with \textit{effective coupling strength} $\Delta$.
This $\Delta$ can vary for different eigenmodes of the ring coupled to the impurity fermions,
but in general, it is proportional to $g$.
Therefore, the effective-coupling Hamiltonian is a tight-binding matrix between some of the eigenmodes of the ring and the impurity fermions.
By the symmetry of the wavefunction, only one zero mode ($\ep = 0$) and one particle-hole pair ($\ep = \pm \alpha J$) of the periodic ring (i.e., zero-flux sector) can hybridize with the impurity (see Methods).
This leads to a simple tight-binding matrix:
\begin{align}
\begin{split}
&\mathcal{H}_{\mathrm{PBC}}^{\rm{eff}}= \bg 0&0&0&\Delta_{1} \\ 0&-\alpha J &0 &\Delta_{2}\\ 0&0&\alpha J &\Delta_2\\ \Delta_1&\Delta_2&\Delta_2&0 \ed\\
\end{split},
\end{align}
where $\Delta_{i} \sim g$ is the effective coupling strength and we assume $\Delta_{i} \equiv \Delta$ for simplicity. This Hamiltonian gives the eigenvalues of the effective-coupling model when $\Delta \ll 1$:
\begin{align}
\ep'_{\rm{PBC}} \approx
 \pm \Delta, \,\,\, \pm\left(\alpha J-\frac{\Delta^2}{\alpha J}\right),
\end{align}
where the prime is added to effective-coupling model eigenvalues $\epsilon^{\prime}$ in order to be distinguished from the hopping-ring model eigenvalues $\epsilon$.
One can simply understand the above results by the perturbation theory. If the eigenmode of the $L = 12$ ring is a zero-energy mode, the effective coupling results in a degenerate perturbation theory with energy correction $\Delta$. On the other hand, if the eigenmodes of the ring have finite energy, the non-degenerate perturbation theory gives rise to second-order corrections. 
Therefore, we can conclude that the ground-state total energy, which is the sum of all negative eigenvalues, has the general expressions (in the unit of $J$):
\begin{align}
E_{\mathrm{PBC}}^{\rm{eff}} \approx 
-A_1\Delta -B_1\Delta^2 -C_1,
\end{align}
where $A_1$, $B_1$, and $C_1$ are positive constants.
In the bound-flux case, there is no zero-energy eigenmode in the plaquette model, so the ground-state energy expression does not include the linear term in $\Delta$ when $\Delta \ll 1$:
\begin{align}
E_{\mathrm{APBC}}^{\rm{eff}} \approx
 -B_2\Delta^2-C_2.
\end{align}
In the effective-coupling model, the ground-state energy difference defined in Eq. (\ref{eq:energy_diff}) translates into $\Delta E = E_{\rm{APBC}}^{\rm{eff}}-E_{\rm{PBC}}^{\rm{eff}}$, which is the measure of the flux-sector transition. It can be modeled as
\begin{align}\label{eq:quadratic}
\Delta E \approx
 A_1 \Delta + (B_1-B_2)\Delta^2 + (C_1-C_2),
\end{align}
 and the coefficients are fitted by the exact-diagonalization result for the 13-site fermion-hopping model, which gives $A_1 \approx 1.163$, $(B_1-B_2)\approx -1.079$, and $(C_1-C_2) \approx -0.261$.
Apparently, one can see that when $\Delta \ll 1$, $\Delta E$ is a concave quadratic function and provides the possibility of first (bound-to-zero) and second (zero-to-bound) flux-sector transitions.
 However, if the predicted second transition happens at the strength $\Delta_2^{*}$ beyond the validity of the quadratic approximation, it may not be seen in the exact diagonalization result. This is exactly what happens for the $S_{\rm{imp}} = 1/2$ case shown in the Fig.~\ref{fig:Majorana_rep}\textbf{c}, where the exact diagonalization curve starts to deviate from the quadratic curve at $\Delta \approx 0.5J$.

The next important question is how to incorporate the higher-spin impurity. Here, we conjecture that the essential ingredient is the additional $\gamma^0_a$ Majorana fermions that provide more entries of the tight-binding matrix. For example, for $S_{\rm{imp}} = 3/2$, one introduces three additional zero modes and considers the perturbative effects on the zero mode ($\epsilon_0 = 0$) and the finite-energy modes ($\epsilon_{\alpha} = \pm \alpha J$) separately: 
\begin{align}
\begin{split}
&\mathcal{H}_{0} = \bg 0 & \Delta & \Delta & \Delta \\ \Delta & 0 & 0 & 0 \\ \Delta & 0 & 0 & 0\\ \Delta & 0 & 0 & 0  \ed, \,\, \epsilon^{\prime}_0 = 0,\pm\sqrt{3}\Delta\\
&\mathcal{H}_{\alpha} = \bg -\alpha J & 0 & \Delta & \Delta & \Delta \\ 0 & \alpha J & \Delta & \Delta & \Delta \\ \Delta & \Delta & 0 & 0 & 0 \\ \Delta & \Delta & 0 & 0 & 0 \\ \Delta & \Delta & 0 & 0 & 0\ed, \,\, \epsilon^{\prime}_{\alpha} \approx 0,\pm\left[\alpha J+\frac{3\Delta^2}{\alpha J}+\mathcal{O}\left(\frac{\Delta^4}{(\alpha J)^3}\right)\right]
\end{split}
\end{align}
This indicates that the higher-spin effect can be incorporated by making the substitution  $\Delta \to \sqrt{2S_{\rm{imp}}}\Delta$. In Fig. \ref{fig:Majorana_rep}\textbf{c}, we show that the higher-spin impurity in the effective-coupling model simply shifts the quadratic curve to the left. This implies that for the higher-spin impurity case,
the first transition point $\Delta_1^{*}$ is smaller,
and the second transition $\Delta_2^{*}$ is more likely to happen when $\Delta \ll 1$ is still valid. This observation from the effective-coupling model, despite its over-simplicity, is consistent with the DMRG results as shown in Fig. \ref{fig:DMRG_results}\textbf{a}. Based on our findings, one may expect that higher-spin impurity cases such as spin-2, 5/2, and so on, also tend to bind the $W_I$ flux under the  $\Delta \to \sqrt{2S_{\rm{imp}}}\Delta$ substitution. However, it is important to note that our effective model does not account for the effect of $\gamma_a^\mu$, which may lead to unexpected results for even larger spins. Therefore, the precise nature of even higher-spin cases requires further numerical evidence to support the stability of the bound-flux sector.

\section*{Discussion}
In this study, we investigated the behavior of $S_{\rm{imp}} = 1$  and $S_{\rm{imp}} = 3/2$  impurities in KSL,
focusing on their effects on the flux-sector transitions and the stability of the bound-flux sector.
Our analysis, using ED and DMRG methods, along with the phenomenological model in the Majorana representation, revealed several key findings: 
First, the local behavior of KSL around a magnetic impurity strongly depends on whether the impurity has a half-integer or integer spin.
This dependence was demonstrated by considering the cases of $S_{\rm{imp}} = 1$ and $S_{\rm{imp}} = 3/2$, and analyzing their impact on flux-sector transitions.
Second, magnetic impurities can bind $Z_2$ fluxes in the lattice, similar to vacancies and quasivacancies.
We observed a phase transition between bound-flux and zero-flux sectors by varying the impurity coupling strength, with the transition point dependent on the impurity's spin magnitude.
Third, for $S_{\rm{imp}} = 3/2$ impurities, a reentrant bound-flux sector was observed, remaining stable under finite magnetic fields. This stability increases with the coupling strength $g$.
Fourth, our ED calculations for fermionic Hamiltonians in the triple-plaquette cluster
revealed the energy differences between the bound-flux and zero-flux states, which are in good agreement with the flux-sector transition points identified in the DMRG study of the spin-basis Hamiltonian. Proper constraints by projection operators are essential for accurate flux-sector transition behavior.

The ability of spin-$S$ impurities to bind $Z_2$ fluxes and its stability against the external magnetic field has practical implications. 
When time-reversal symmetry is broken, flux binding to the impurity site results in the formation of localized Majorana zero modes, which are essential for realizing Ising anyons. These anyons exhibit non-Abelian statistics, making them valuable for topological quantum computation. 

Compared to an Ising anyon bound at a vacancy \cite{Willans2010}, the anyon found at the magnetic impurity site offers a more advantageous way to access low-energy Majorana-bound states through its magnetic channel. This unique feature of entangled Ising anyons can be observed in the dynamical correlation function of impurity spins, especially by focusing on the low-energy spectra.

Spectroscopy techniques with high spatial resolution, such as STM or nitrogen-vacancy (NV) center magnetometry, could further advance our understanding of the ground-state flux sector in Kitaev spin liquid phases. The low-energy spectrum carries crucial information about the ground-state flux sector, and both STM and NV center magnetometry are capable of probing these spectral features. STM might reveal the dynamical correlation function of impurity spins through spin-dependent tunneling \cite{Takahashi2023, KaoPRL2024, KaoPRB2024, Bauer2024}, while NV center magnetometry could detect magnetic noise linked to the same correlation function \cite{Casola_NVcenter_review} as well as emergent gauge field \cite{PALee2023_NVcenter}. Observations of low-energy features, including Majorana zero modes, could potentially inform efforts toward realizing topological qubits composed of Ising anyons.

In addition, the  even-odd effect on magnetization is another potential signature worth investigating. 
 Local measurements of spin moments, such as those achievable via NV center magnetometry in the absence of a magnetic field, could provide clear evidence of this effect for a magnetic impurity embedded in Kitaev spin liquid phases. However, the spin Hamiltonian discussed in this work may be too simplified to accurately capture magnetic properties in real candidate materials \cite{Lee2023}.
  Furthermore, introducing more than two magnetic impurities into the system could result in an RKKY-like interaction mediated by itinerant Majorana fermions in the bulk \cite{Legg2019,Dhochak2010}.
  While this interaction could obscure or destabilize the even-odd effect, it may simultaneously promote magnetic ordering of impurity magnetic moments, even within an otherwise spin-liquid phase.

\section*{Methods}
\subsection*{Implementation of DMRG}
All DMRG calculations were performed using the NVIDIA Data Center GPU R470 Driver with the ITensorsGPU.jl package \cite{ITensorsGPU}.
To ensure qualitative accuracy, determined by the expectation value of all plaquettes for both ${W}_p$ in the bulk and ${W}_I$ (plus, even ${\mathbb{W}}_a$ for a half-integer impurity case) at impurity sites, we required an adequate bond dimension $d$ depending on the impurity size and a good energy tolerance $\delta E \leq 1\times 10^{-7}$ while satisfying the cutoff at each sweep to be $\leq 1\times 10^{-9}$.
For instance, in the single spin-3/2 impurity problem, we set the maximum bond dimension $d_{\rm{max}}$	to 3000 to ensure the cutoff condition.
Additionally, we performed DMRG calculations five times independently for each parameter point and selected the result with the lowest ground state energy realizing reasonable flux expectation values.

\subsection*{Construction of projection operators for higher-spin}
Here, we first review the Majorana representation for arbitrary spin size introduced by Ma \cite{Ma2023},  and then construct projection operators applied in the ED calculation of the many-body Majorana Hamiltonian for the $S_{\rm{imp}}=3/2$ case.
The starting point is to consider a spin-$\tilde{S}$ operator as $2\tilde{S}$ of spin-1/2s:
\begin{align}
{\tilde{S}}^\mu=\sum_{a=1}^{2\tilde{S}}{S}_a^\mu =\sum_{a=1}^{2\tilde{S}}\frac{\sigma_a^\mu}{2},
\end{align}
where ${S}^\mu$ denotes the $\mu$-component of spin-1/2 operator with $\bm{\sigma}$ being Pauli matrices,
and $a$ indicates the flavor degree of freedom.
Then, the Majorana representation is applied for each spin-1/2 as
${\sigma}_a^\mu=i\gamma_a^\mu \gamma_a^0$
\footnote{
 The order of Majorana operators is slightly modified from the original to follow Kitaev's Majorana representation \cite{Kitaev2006}.}.
Here four Majorana fermions $\gamma^x,\gamma^y,\gamma^z$ and $\gamma^0$  are introduced for each flavor.
As a result, 
\begin{align}
{\tilde{S}}^\mu=\frac{1}{2}\sum_{a=1}^{2\tilde{S}}i\gamma_{a}^\mu\gamma_a^0.
\end{align}
Note that for spin-1/2 operators, we use the usual symbol of Majorana fermions $b^x,b^y,b^z$ and $c$ instead of $\gamma_a^\eta$ ($\eta=x,y,z,0$).
The local $Z_2$ gauge field operators $u_{jk}^\mu=ib_j^\mu b_k^\mu$,
which connect two spin-1/2s at sites $j$ and $k$ on the $\mu$-bond, commute with the Hamiltonian $H$.

Since the Hilbert space is expanded in this representation— both for the bulk spin-1/2s and impurity sites—we need two  kinds of projection operators to ensure the correct physical states are selected.
 The first one is required to ensure the commutation relations of Pauli matrices at each spin-1/2 or flavor in the Majorana representation, denoted by the condition $C_\gamma^a: {D}_a=\gamma_a^x\gamma_a^y\gamma_a^z\gamma_a^0=1$ in Ref. \cite{Ma2023}.
Resulting operator is
\begin{align}
{P}_a = \frac{1+{D}_a}{2}\qquad(a=1,2,\cdots,2\tilde{S}).
\end{align}
Note that ${P}_a$ satisfies the condition of a projection operator automatically as $({P}_a)^2={P}_a$.

The second projection operator,  which is required only for higher-spin ($\tilde{S}>1/2$) cases, mixes different flavors to ensure $|\tilde{S}^2|=\tilde{S}(\tilde{S}+1)$.  This condition is represented as $C_s: \sum_\mu (\sum_{a=1}^{2\tilde{S}} \gamma_a^\mu\gamma_a^0)^2=-4\tilde{S}(\tilde{S}+1)$ in Ref. \cite{Ma2023}.
For $\tilde{S}=3/2$, corresponding operator can be represented as 
\begin{align}
{P}_{\tilde{S}=3/2}=-\frac{1}{6}\left[-3 + \sum_{\mu}\sum_{a>b}\gamma_a^\mu\gamma_a^0\gamma_b^\mu\gamma_b^0\right].
\end{align}
It is worth mentioning that ${P}_a$ is necessary for ${P}_{\tilde{S}=3/2}$ to satisfy the condition of projection operator since
\begin{align}
   ({P}_{\tilde{S}=3/2})^2 = -\frac16\left[-3 + \sum_{\mu}\sum_{a>b}\,\frac{4-{D}_a{D}_b}{3}\,\gamma_a^\mu\gamma_a^0\gamma_b^\mu\gamma_b^0\right].
\end{align}
In the  Majorana Hamiltonian with four-body interactions in the vicinity of the spin-$3/2$ impurity,
these two kinds of projection operators $P_{\tilde{S}=3/2}$ and $P_a$ for $a = 1,2,3$ are essential for accurately calculating the energy difference between the two flux sectors.

\subsection*{Eigenenergies and eigenfunctions of the periodic ring}
\begin{figure*}
    \centering
    \includegraphics[width=170mm]{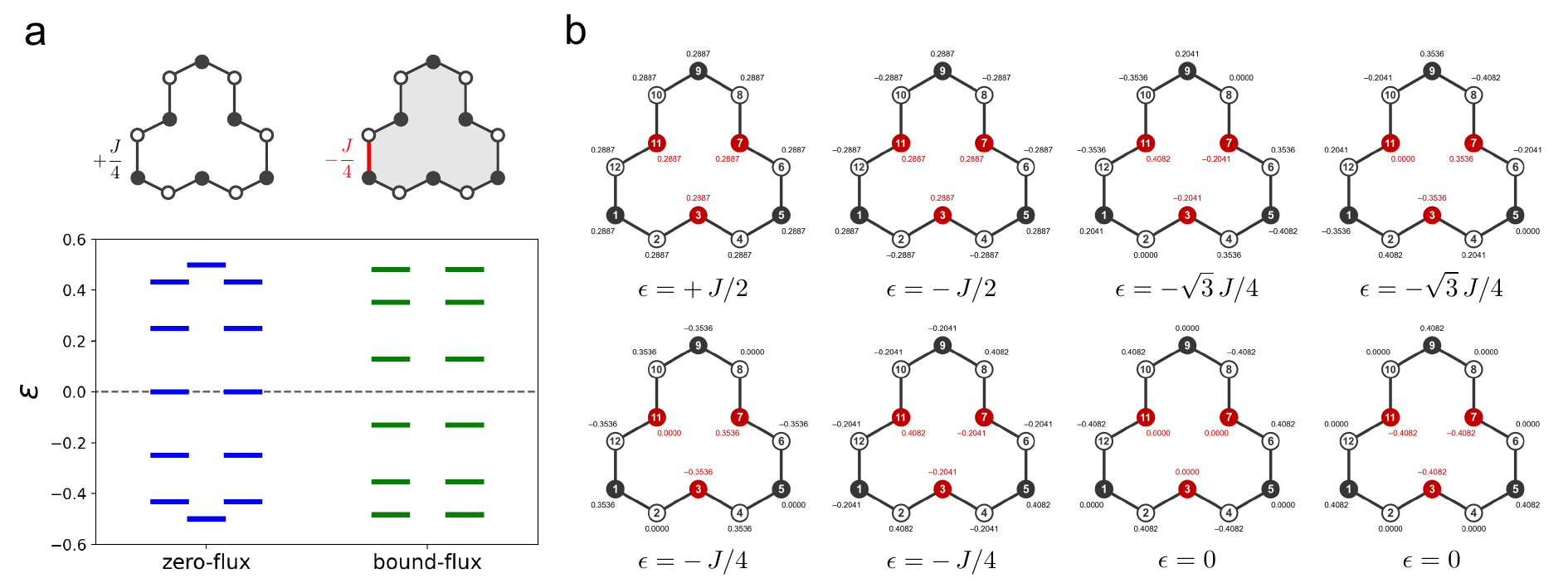}
    \caption{  Eigenstates of the $L = 12$ fermionic ring.
    \textbf{a} The energy spectrum of the fermionic ring with periodic or antiperiodic boundary. The former corresponds to the zero-flux sector of $W_I$, while the latter corresponds to the bound-flux sector.
    \textbf{b} The eigenfunctions for some selected eigenmodes. The sites in red can directly hop to the additional impurity fermion. However, if the sum over amplitudes on the red sites vanishes, the effective coupling will be zero.}
    \label{fig:periodic_ring}
\end{figure*}

Here we provide details of the $L = 12$ periodic and antiperiodic rings used in the effective-coupling model. The energy-level spectrum calculated by Eq. \eqref{eq:epsilon_ring} is shown in Fig. \ref{fig:periodic_ring}\textbf{a}. This spectrum can be verified by the diagonalization of the tight-binding model as well. The periodic ring, which corresponds to the zero-flux sector, contains two zero-energy modes. This is the crucial difference from the antiperiodic ring, which corresponds to the bound-flux sector. It is important to note that not all eigenmodes can couple to the impurity. First, the impurity fermion only directly hops to three nearest-neighbor sites, which are colored in red in Fig.~\ref{fig:periodic_ring}\textbf{b}. Second, one can straightforwardly symmetrize the eigenmodes for each degenerate pair of modes, such that only some of the modes couple to the impurity fermion.  Specifically, the 12-site model on a periodic ring has a $C_{3v}$ point-group symmetry, such that only three modes are subject to the transformation of the trivial irreducible representation and couple to the impurity site. Because the hopping integral is based on the overlap of one site (impurity) and the three neighboring sites simultaneously, one can simply assume $\psi_{\rm{imp}} = 1$ as the impurity wavefunction and the sum of the amplitude of the three sites determines the effective coupling. Specifically, for a given eigenmode of the fermionic ring $\psi_{\rm{ring}}$, if
\begin{align}
\Delta \sim \sum_{j}\psi_{\rm{imp}}\,g\,\psi_{{\rm{ring}},j} = \sum_{j}g\,\psi_{{\rm{ring}},j} = 0
\end{align} with $j \in (3, 7, 11)$ in Fig.~\ref{fig:periodic_ring}\textbf{b}, the eigenmode is simply decoupled to the impurity site.
From Fig.~\ref{fig:periodic_ring}\textbf{b}, we see that only the eigenmodes with $\epsilon = \pm J/2$ and one of the zero modes $\epsilon = 0$ can couple to the impurity. This is numerically verified by the exact diagonalization of the 13-site tight-binding matrix. A similar analysis can be made on the antiperiodic ring for the bound-flux sector. However, since there is no zero-energy mode in the spectrum, the perturbative effect on the eigenmodes by the impurity fermion starts from the second-order correction. This leads to the qualitative argument based on our effective-coupling model, which is discussed in the main text.

\section*{Data availability}
The datasets generated during and/or analyzed during the current study are available from the corresponding author upon reasonable request.

\section*{Code availability}
The codes used during the current study are available from the corresponding author upon reasonable request.

\section*{Acknowledgments}
The authors thank Han Ma, G\'abor Hal\'asz\, and Bo Xiao for fruitful discussions.
M.O.T. is supported by a Japan Society for the Promotion of Science (JSPS) Fellowship for Young Scientists
and by the Program for Leading Graduate Schools: “Interactive Materials Science Cadet Program.”
This work is supported by JST CREST Grant No. JPMJCR19T5, as well as JSPS KAKENHI No. JP22J20066 and No. JP23K20828. W.H.K.  and N.B.P. were supported
by the U.S. Department of Energy, Office of Science, Basic Energy Sciences under Award No. DE-SC0018056.    N.B.P. also acknowledges the hospitality of the Aspen Center for Physics.

\section*{Author Contributions}
M.O.T. and N.B.P. devised the project.
M.O.T. performed numerical calculations for the spin system and analyzed them using the Majorana fermion representation.
W.H.K. proposed and carried out the effective-coupling model calculations. 
M.O.T., W.H.K, S.F., and N.B.P. contributed to the interpretation of the results and the writing of the paper.

\section*{Competing interests}
The authors declare no competing interests.

\bibliography{main}
\end{document}